\documentclass[prb,aps,twocolumn,amstex,preprintnumbers,amsmath,amssymb,array,tabularx]{revtex4}
\usepackage{graphicx}% Include figure files
\usepackage{dcolumn}% Align table columns on decimal point
\usepackage{bm}% bold math
\usepackage{lscape} %lanscape/portrait
\usepackage{amsmath}
\usepackage{amsxtra}
\usepackage{tabularx}

\DeclareGraphicsExtensions{.jpg,.pdf,.eps}

\begin{document}
\title{Edge effects in finite elongated carbon nanotubes}

\author{$\mbox{Oded Hod}$, $\mbox{Juan E. Peralta}$, and $\mbox{Gustavo E. Scuseria}$}

\affiliation{Department of Chemistry, Rice University, Houston, Texas
77005}

\date{\today}

\begin{abstract}
The importance of finite-size effects for the electronic structure of
long zigzag and armchair carbon nanotubes is studied. We analyze the
electronic structure of capped (6,6), (8,0), and (9,0) single walled
carbon nanotubes as a function of their length up to $60$~nm, using a
divide and conquer density functional theory approach.  For the
metallic nanotubes studied, most of the physical features appearing in
the density of states of an infinite carbon nanotube are recovered at
a length of $40$~nm.  The (8,0) semi-conducting nanotube studied
exhibits pronounced edge effects within the energy gap that scale as
the inverse of the length of the nanotube.  As a result, the energy
gap reduces from the value of $\sim 1$~eV calculated for the periodic
system to a value of $\sim 0.25$~eV calculated for a capped $62$~nm
long CNT.  These edge effects are expected to become negligible only
at tube lengths exceeding $6$~$\mu$m.  Our results indicate that
careful tailoring of the nature of the system {\em and} its capping
units should be applied when designing new nanoelectronic devices
based on carbon nanotubes.  These conclusions are expected to hold for
other one-dimensional systems such as graphene nanoribbons, conducting
polymers, and DNA molecules.
\end{abstract}

\maketitle
%\newpage 
%%%%%%%%%%%%%%%%%%%%%%%%%%%%%%%%%%%%%%%%%%%%%%%%%%%%%%%%%%%%%%%%%%%%%%%%%%%
%%%%%%%                       BODY OF TEXT                          %%%%%%%
%%%%%%%%%%%%%%%%%%%%%%%%%%%%%%%%%%%%%%%%%%%%%%%%%%%%%%%%%%%%%%%%%%%%%%%%%%%

%%%%%%%%%%%%%%%%%%%%%%%%%%%%%%%%%%%%%%%%%%%%%%%%%%%%%%%%%%%%%%%%%%%%%%%%%%%

Carbon nanotubes (CNTs) have been suggested as potential candidates
for replacing electronic components and interconnects in future
nanoelectronic devices.~\cite{Baughman2002} Experiments have revealed
the possibility of obtaining a wide range of electronic behavior when
studying these systems, ranging from coherent
transport~\cite{Frank1998,Liang2001} suitable for interconnects, to
field effect switching
capabilities~\cite{Tans1998,Postma2001,Bachtold2001,Javey2003} needed
for electronic components design.  While earlier conductance
experiments~\cite{Tans1997, Tans1998, Frank1998, Yao1999, Postma2001,
Bachtold2001, Liang2001, Javey2003} involved long segments of CNTs,
new technologies allow for the fabrication of ultrashort CNT junctions
where reproducible room temperature ballistic transport at the high
bias regime is observed.~\cite{Javey2004-1,Javey2004-2} Nevertheless,
the reduced length of the CNT junctions, which is essential to
suppress electron-phonon scattering, introduces important physical
phenomena such as quantum confinement and edge effects.  Several
theoretical studies have emphasized the importance of such effects
when considering the electronic structure,~\cite{Wu2000,Li2005,
Chen2006} electric transport,~\cite{Anantram1998, Orlikowski2001,
Compernolle2003, Jiang2005, Nemec2006} and magnetic~\cite{Okada2003,
Chen2004-2, Chen2005} properties of finite CNTs.  These effects are
expected to be manifested in experiments involving the dielectric
screening constants,~\cite{Lu2004} optical
excitations~\cite{Chen2004-1} and the Raman
spectroscopy~\cite{Saito1999} of such systems.  Since it is predicted
that the physical characteristics of {\em finite} CNTs are
considerably different from those of their {\em infinite}
counterparts, it is essential to identify the limit at which finite
size effects have to be taken into account in order to properly
describe their electronic properties.

Therefore, a fundamental question arises: Is there a universal CNT
length below which finite size effects become important?
\input{epsf}
\begin{figure}[h]
  \begin{center}
    $\begin{array}{r c l c c}
      \epsfysize=1.30cm
      \epsffile{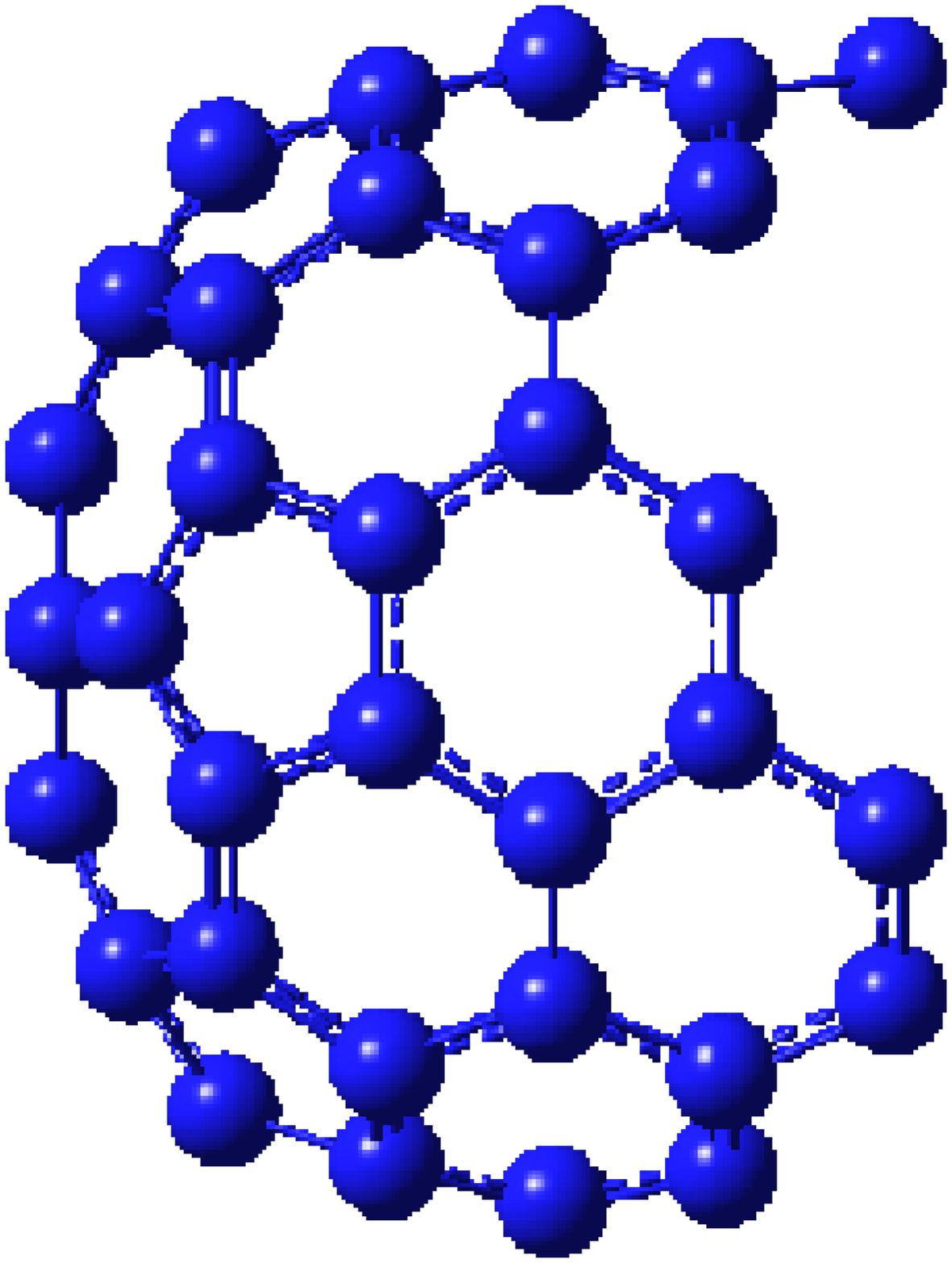}   &
      \epsfysize=1.30cm
      \epsffile{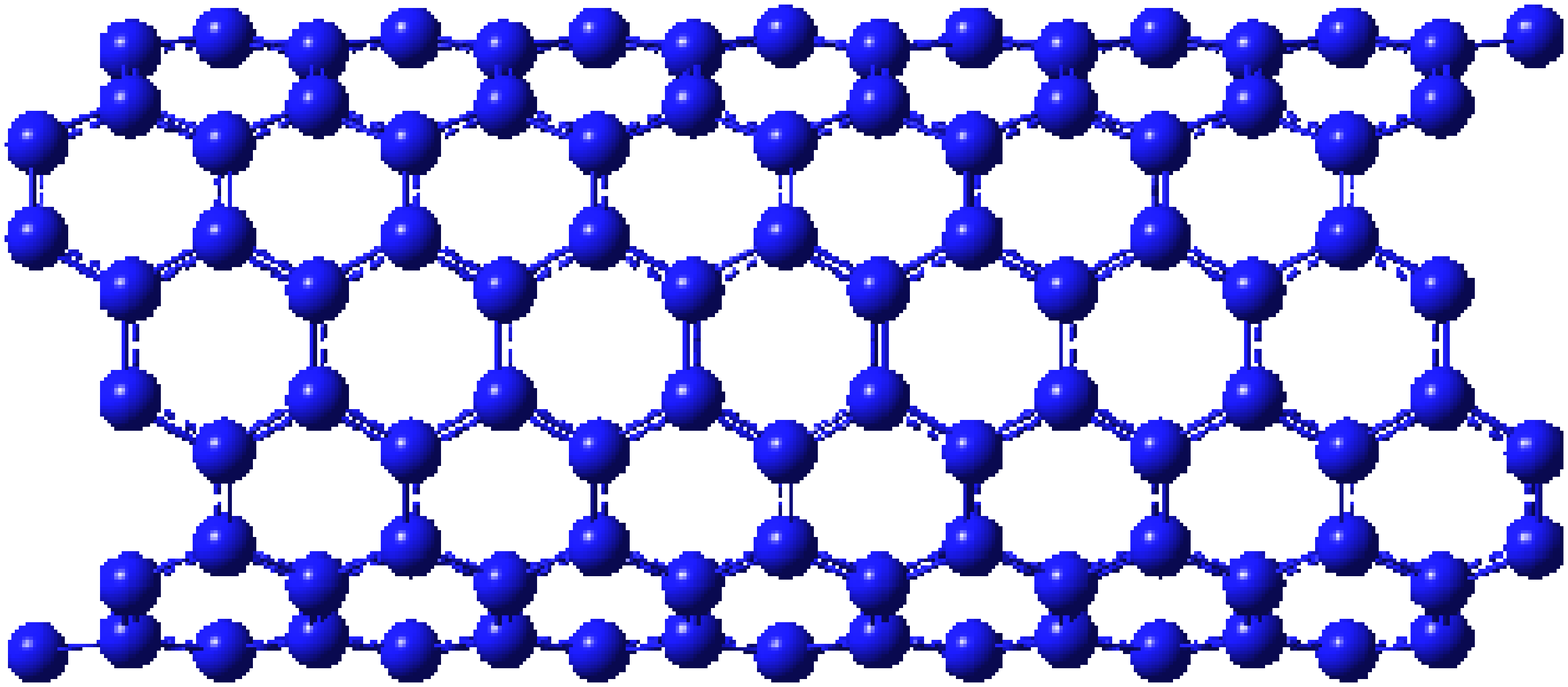} &
      \epsfysize=1.30cm
      \epsffile{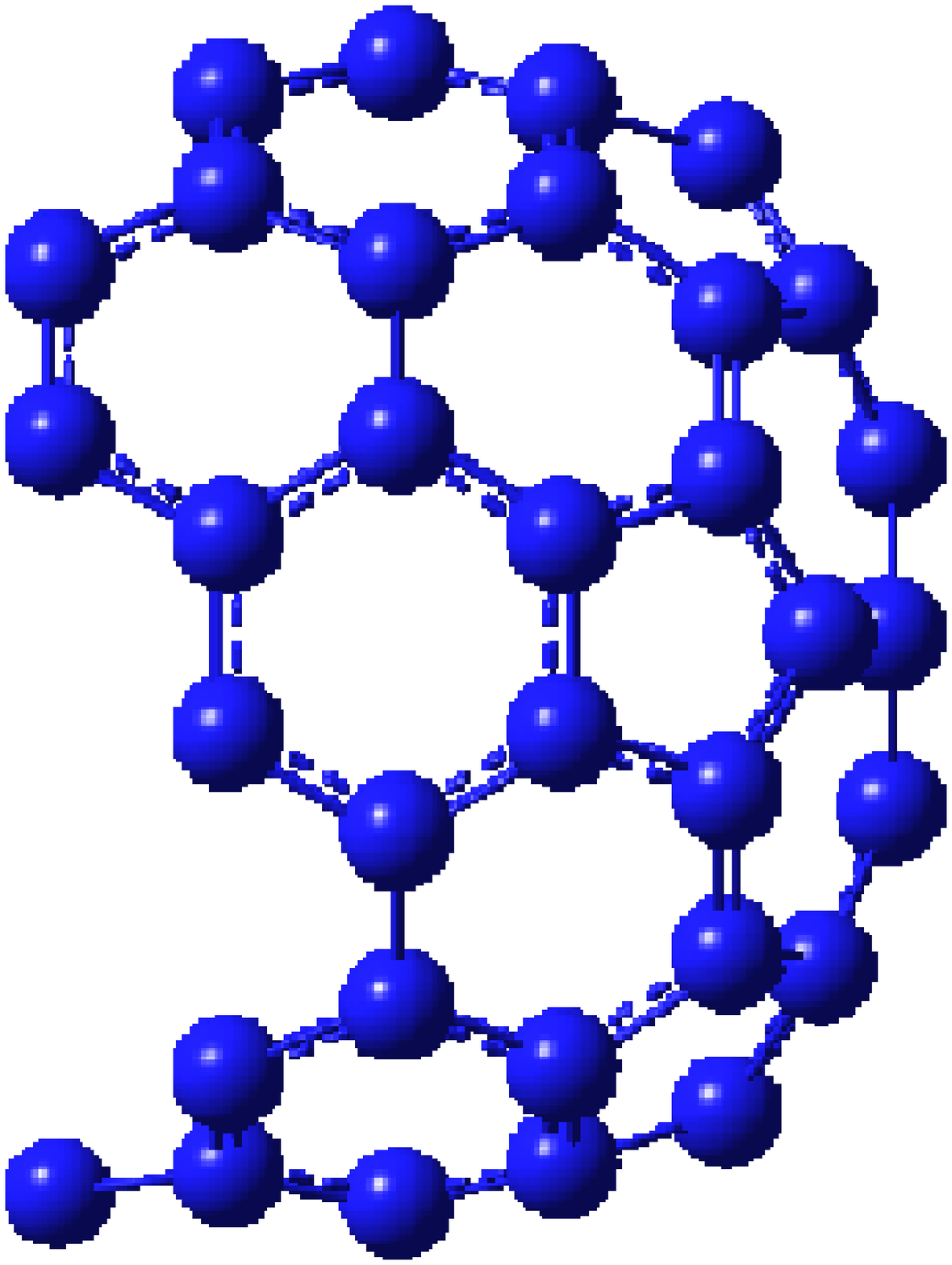}   & &
      \epsfysize=1.30cm                   
      \epsffile{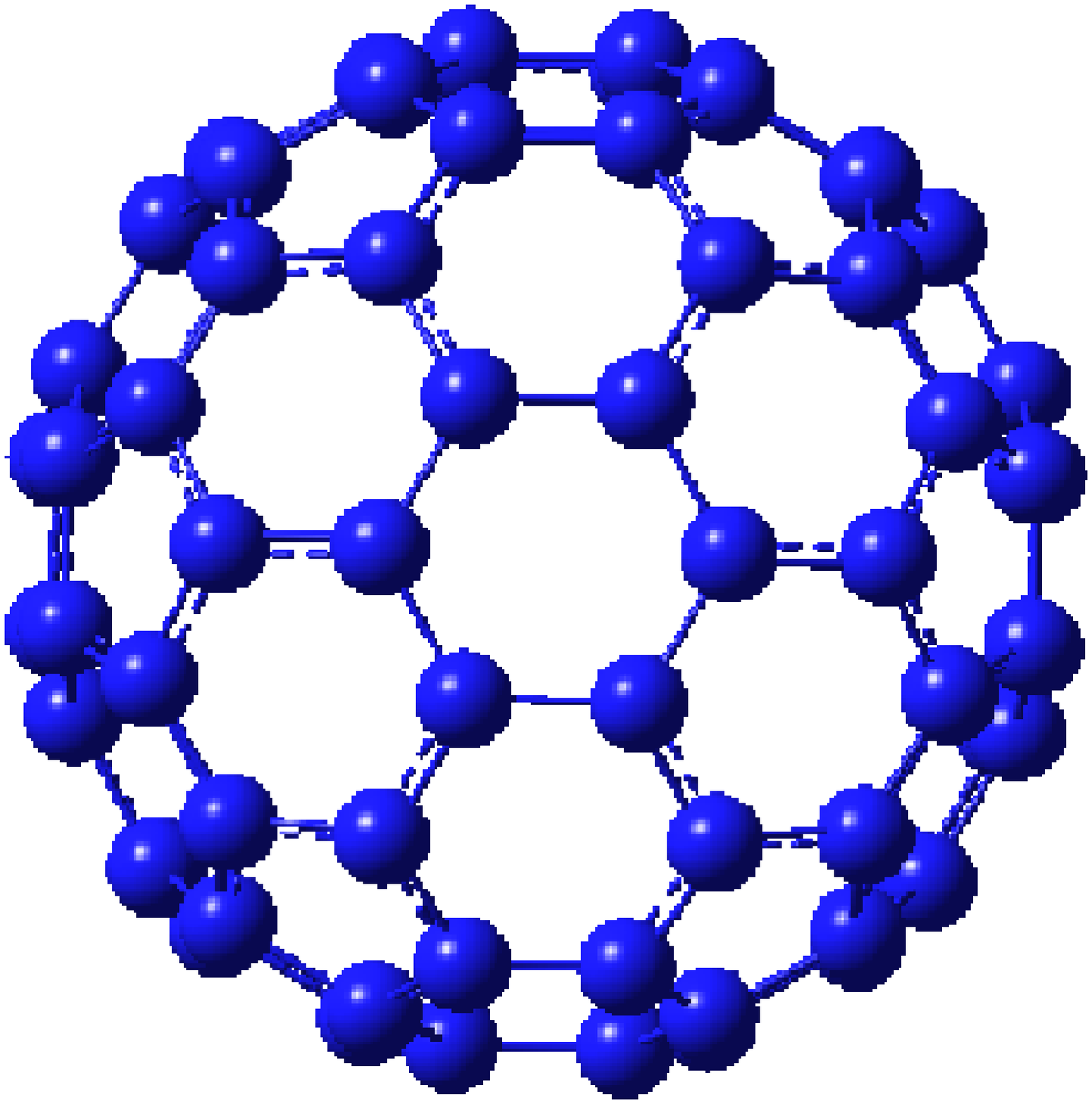} \\ \\ \\
      \epsfysize=1.30cm
      \epsffile{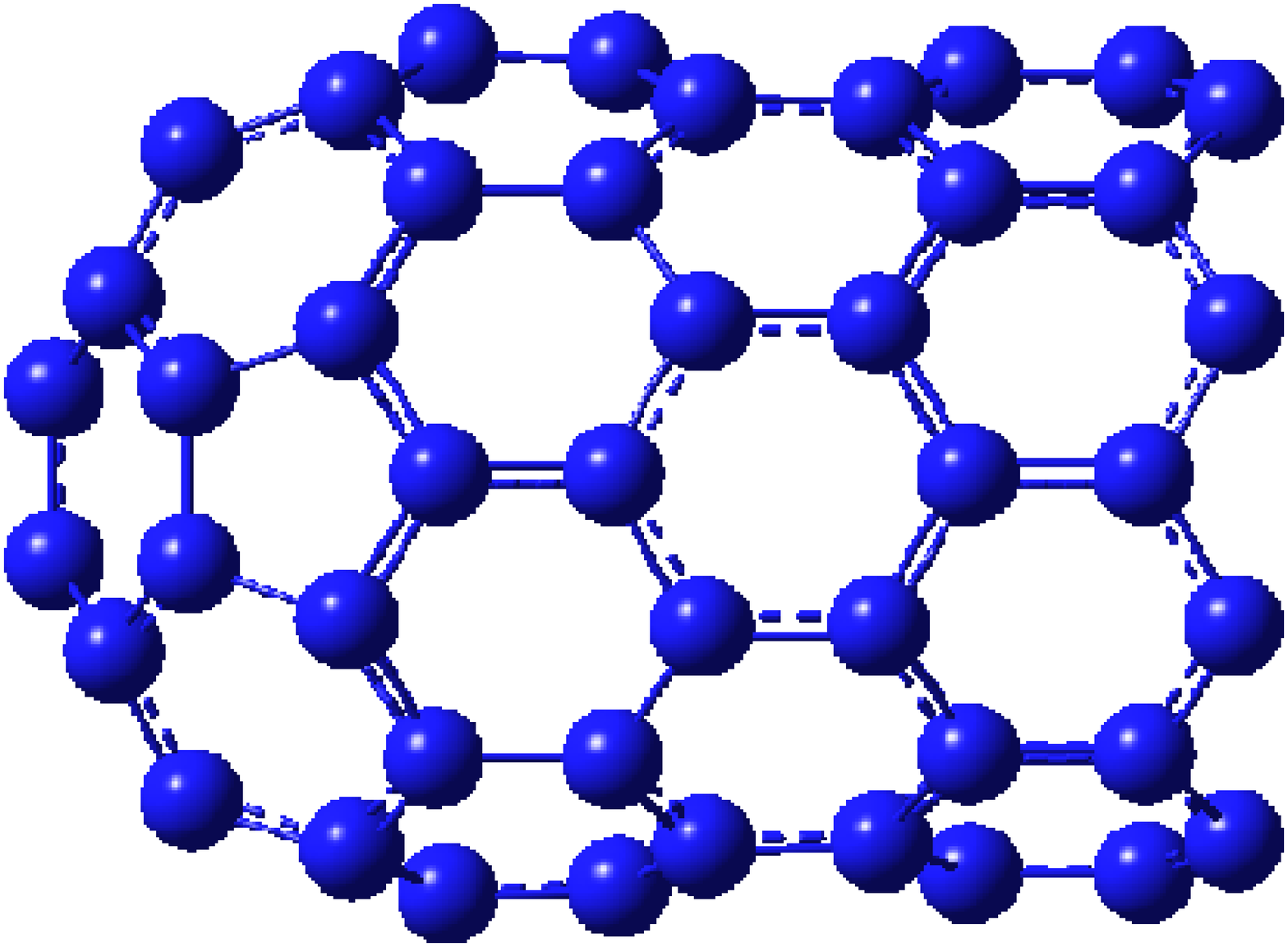}   &
      \epsfysize=1.30cm
      \epsffile{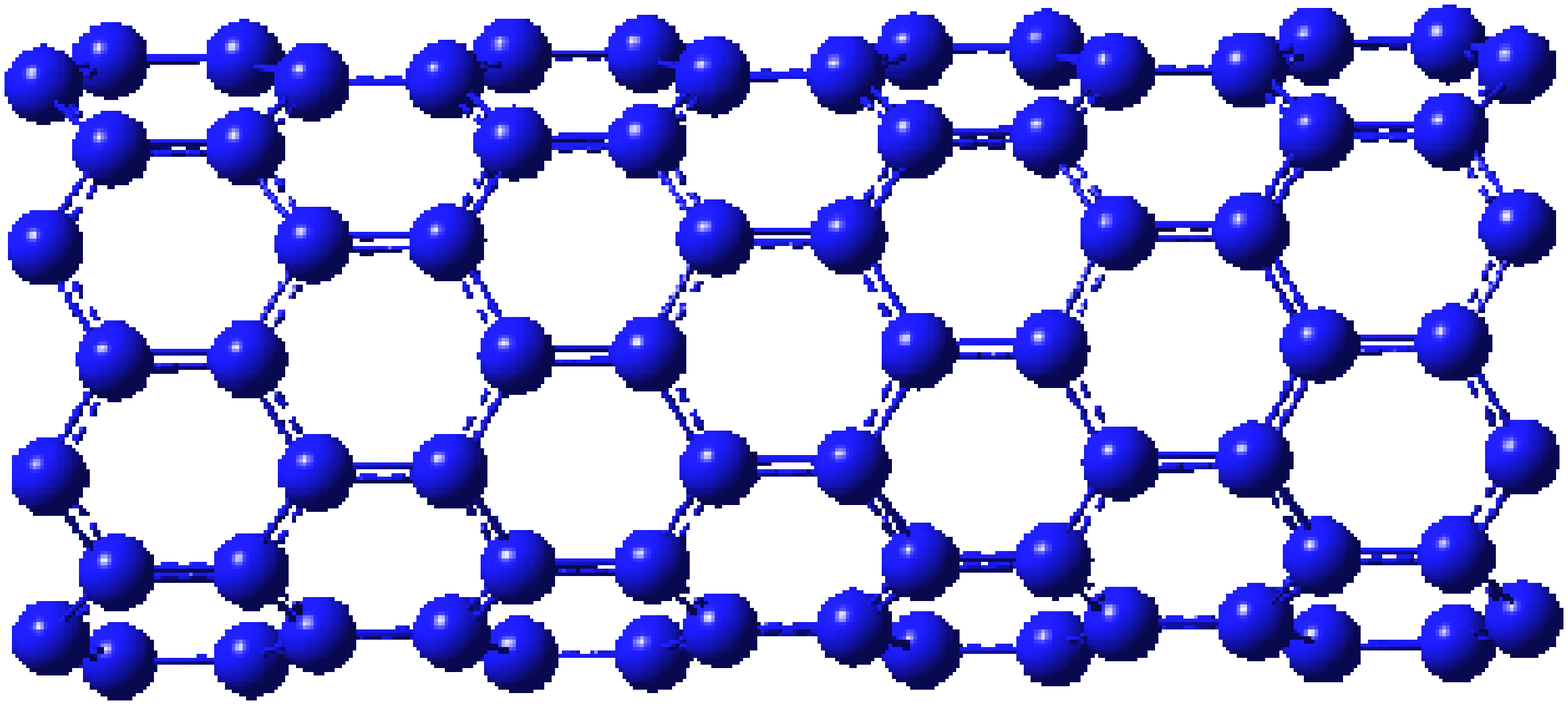}   &
      \epsfysize=1.30cm
      \epsffile{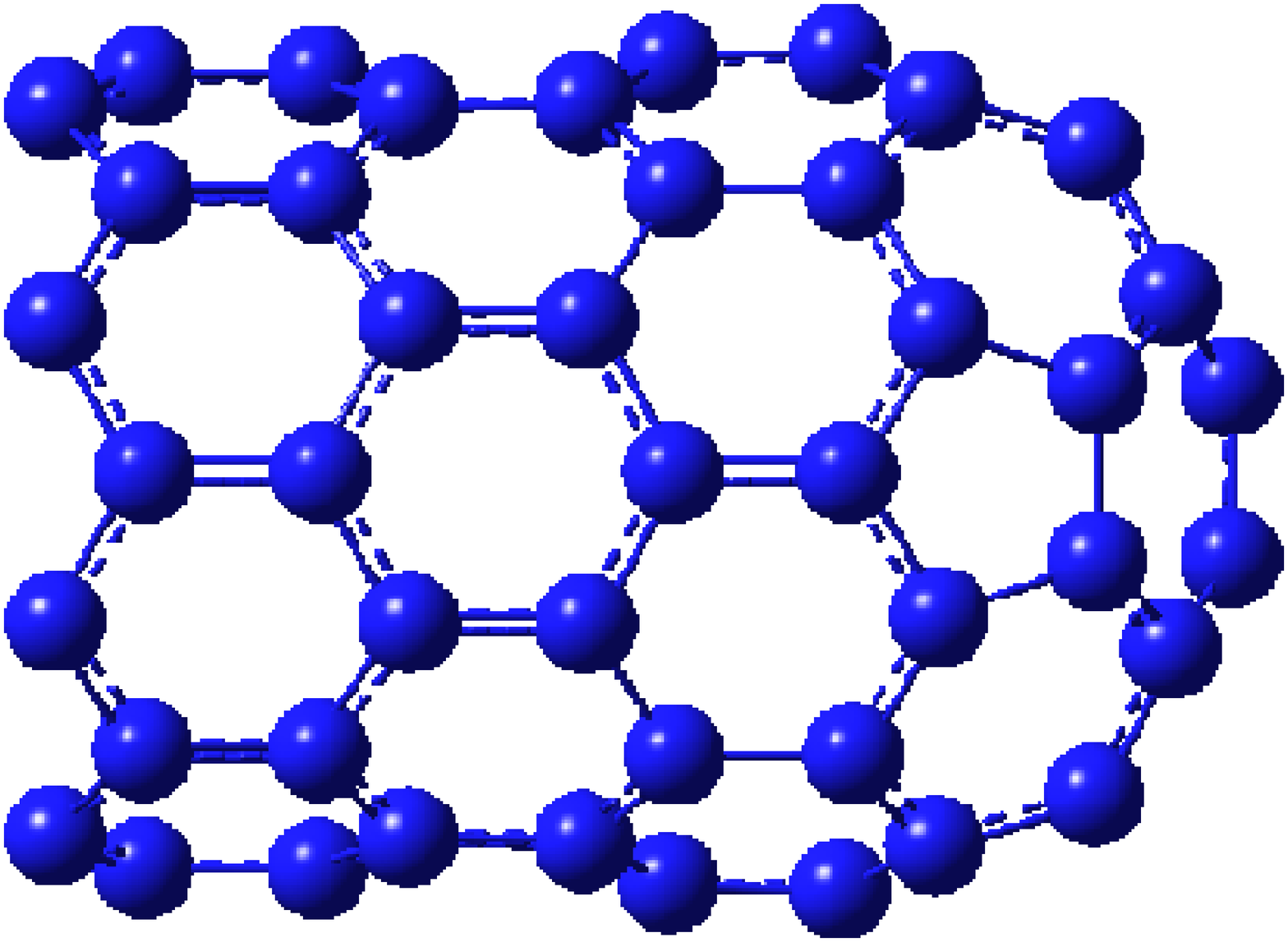}   & &
      \epsfysize=1.30cm
      \epsffile{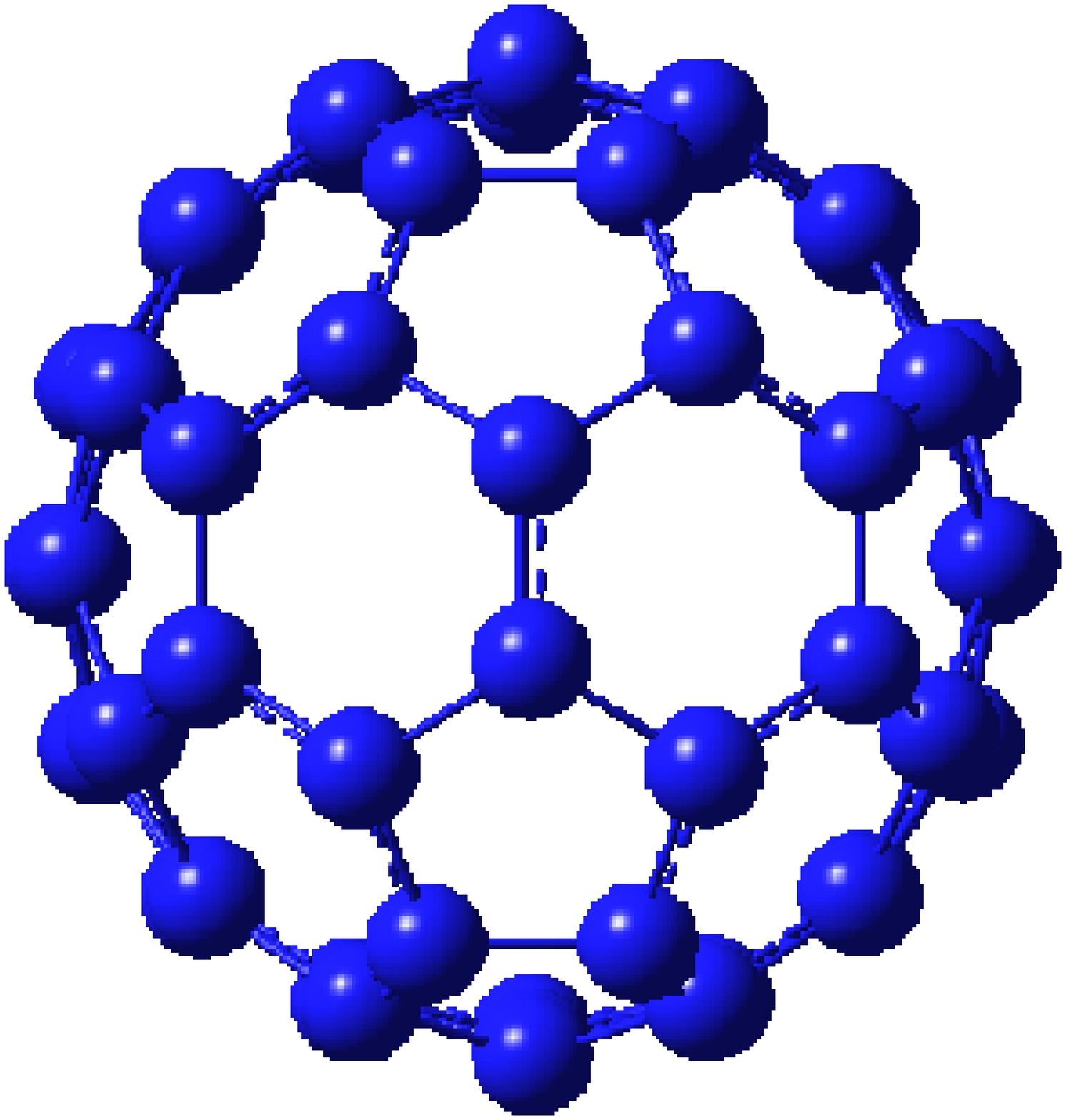} \\ \\ \\
      \epsfysize=1.30cm
      \epsffile{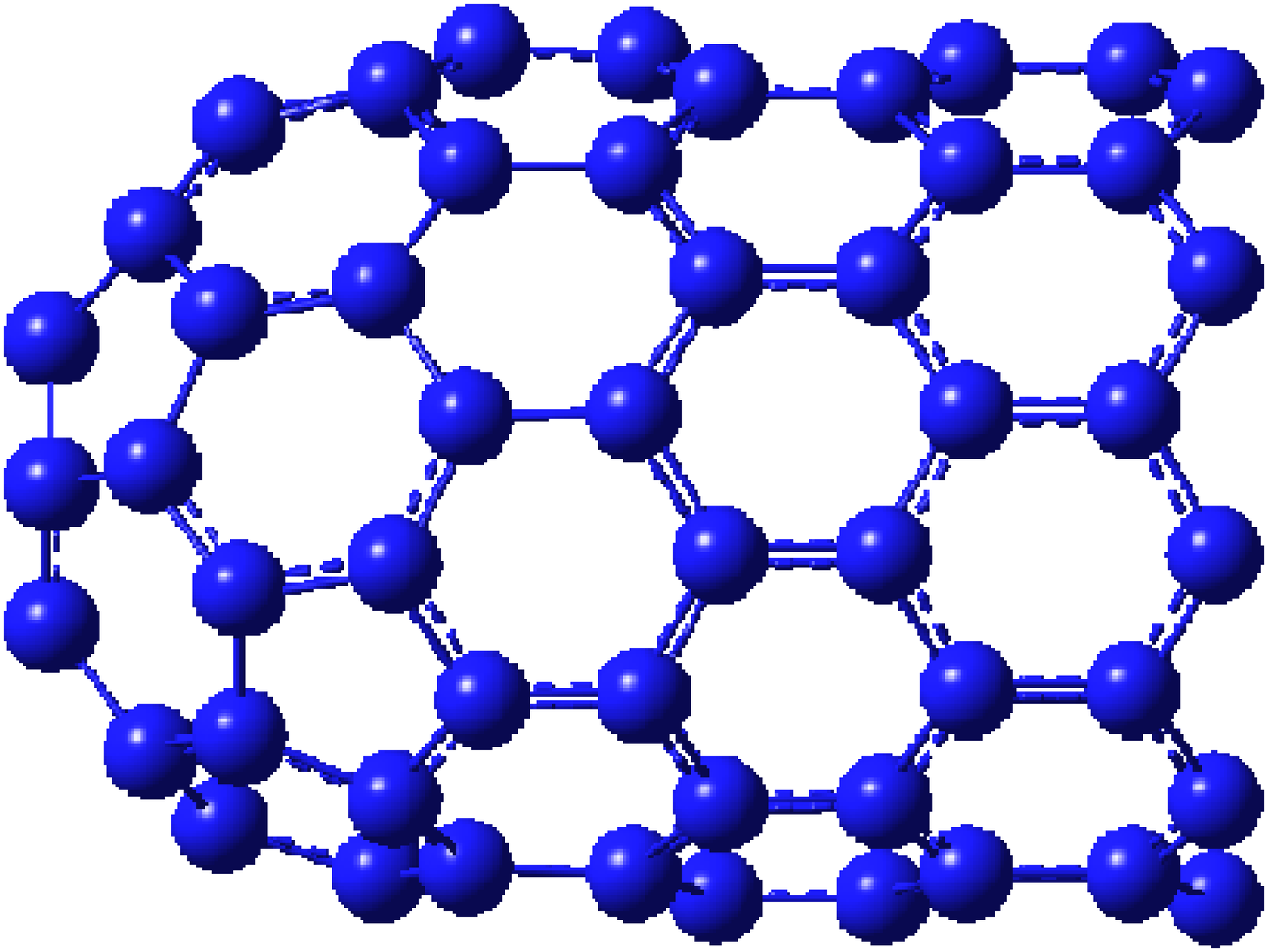}   &
      \epsfysize=1.30cm
      \epsffile{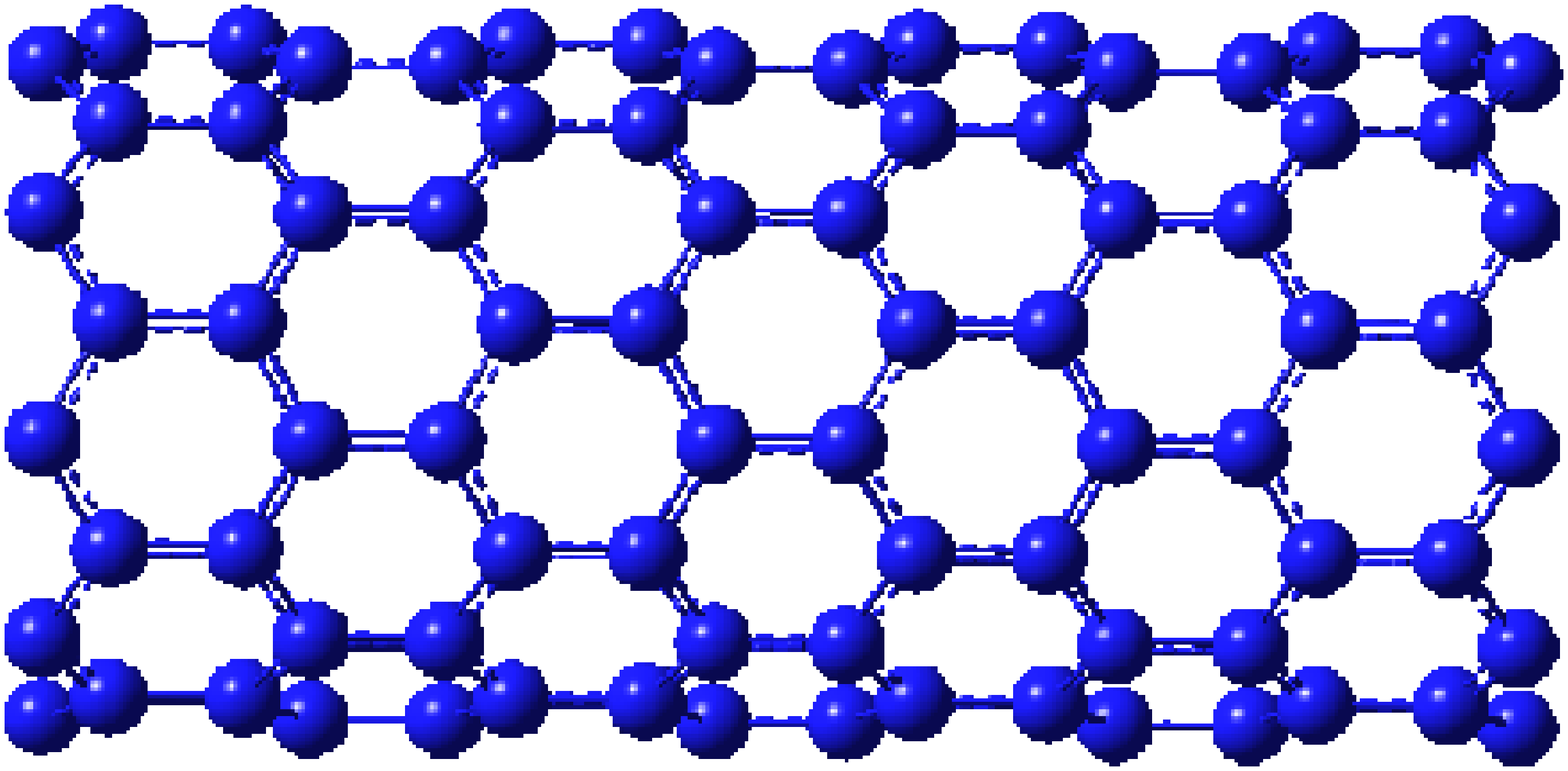}   &
      \epsfysize=1.30cm
      \epsffile{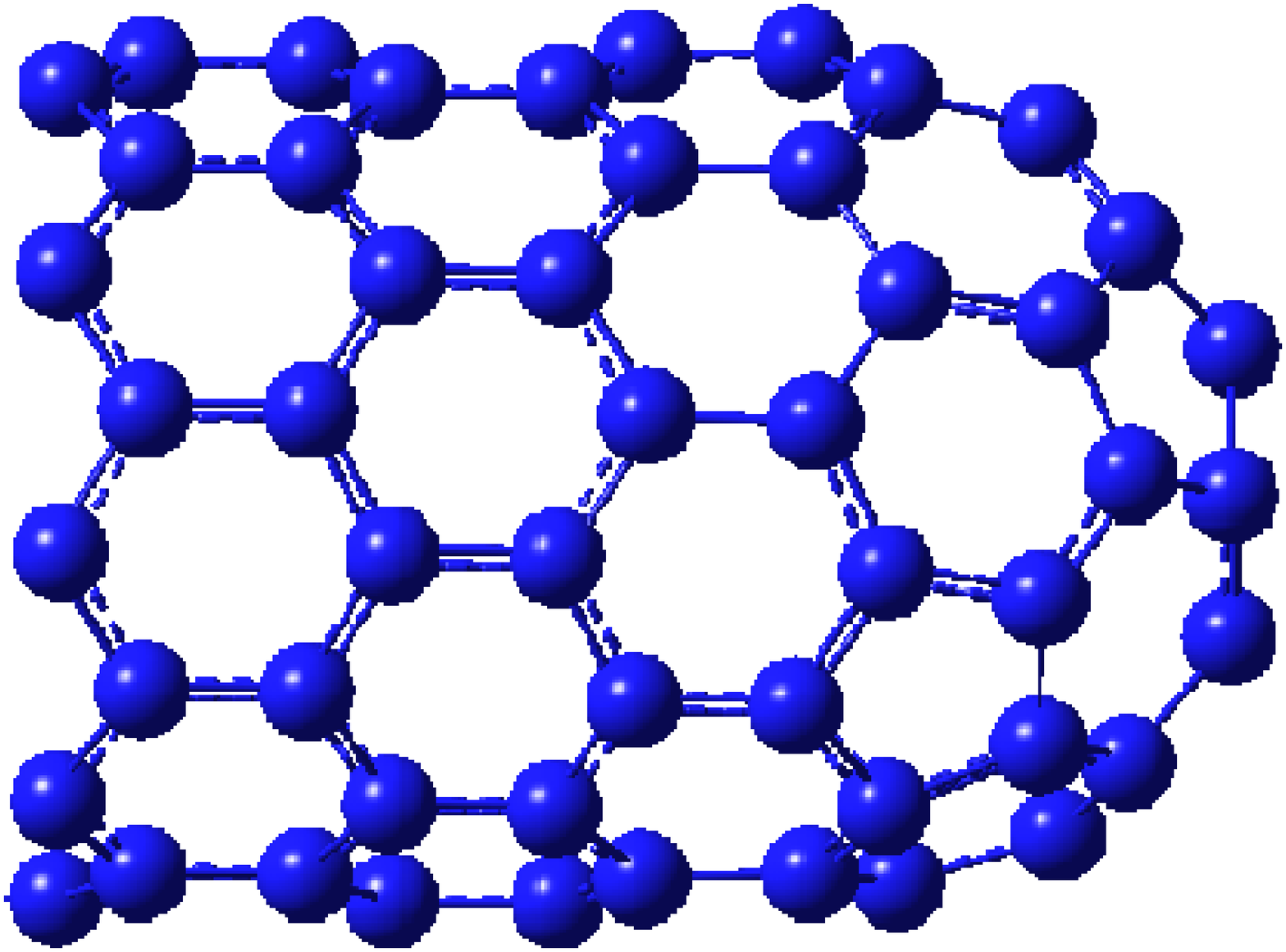}   & &
      \epsfysize=1.30cm
      \epsffile{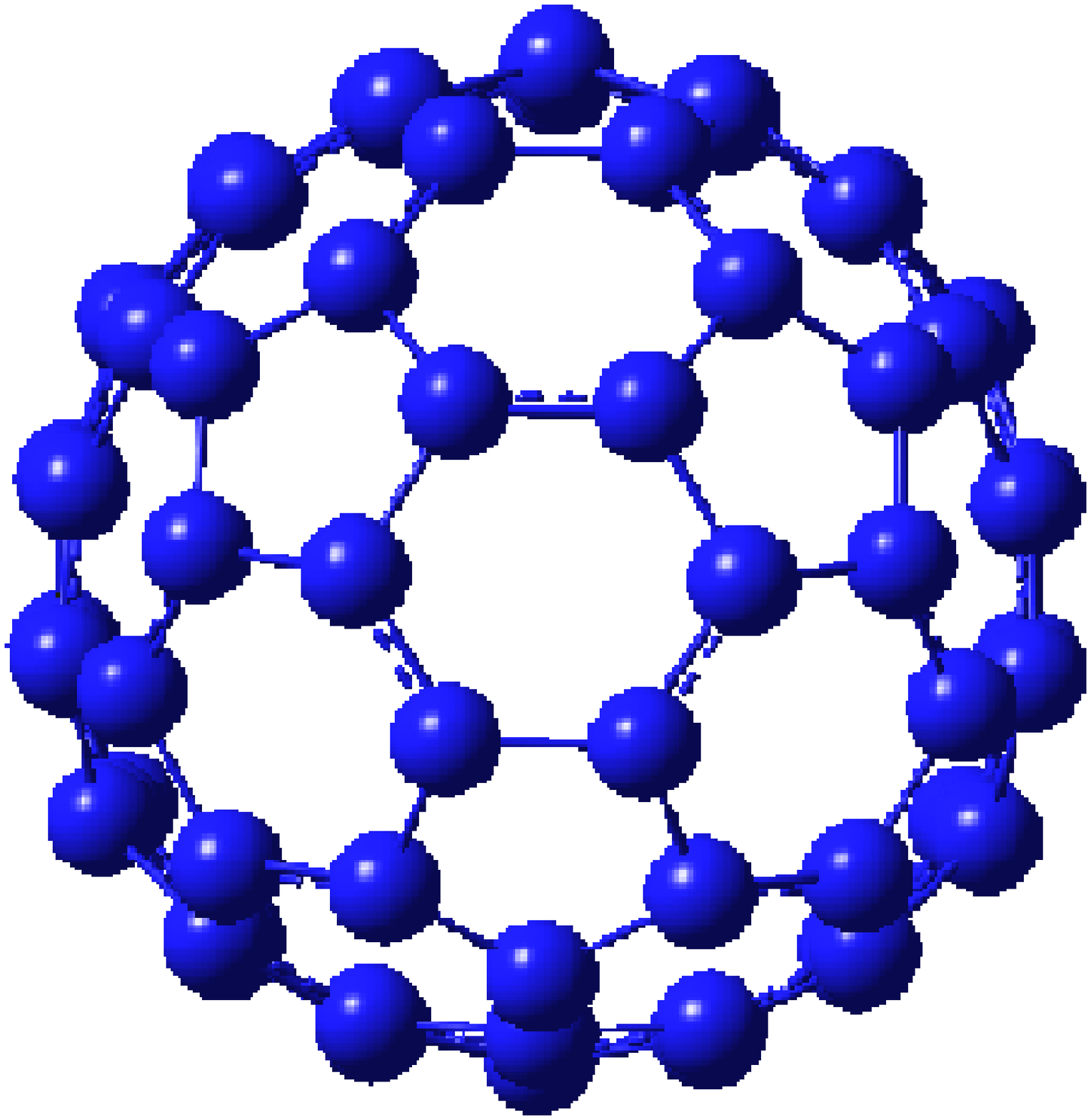}
    \end{array}$
  \end{center}
  \caption{Optimized geometries of a (6,6) (upper panel), (8,0)
    (middle panel), and (9,0) (lower panel) CNTs.  Shown are side
    views of the left capping end, right capping end, and the central
    part which can be replicated to produce the finite elongated
    system.  An axial view of one of the caps of each CNT is
    depicted on the right.}
  \label{Fig: CNT Geometries}
\end{figure}
The aim of this Letter is to answer this question.  To this end, we
present a novel density functional theory (DFT) study of the
electronic structure of CNTs as a function of their length, up to
$62$~nm.  Using a divide and conquer approach for first-principles
electronic structure and transport calculations through finite
elongated systems,~\cite{Hod2006-2} we study the finite size effects
on the electronic structure of capped (6,6), (8,0), and (9,0) CNTS.  A
careful comparison with the electronic structure of infinitely long
periodic CNTs enables us to determine the limit at which a finite CNT
can be fairly approximated by its infinite periodic counterpart.

Our results show that for the metallic CNTs studied herein, most of
the physical features appearing in the density of states (DOS) of the
infinite periodic system are recovered at a length of $40$~nm.  A
similar picture arises for the semi-conducting CNT considered.
However, a subtle difference exists.  Pronounced features in the DOS
resulting from edge effects appear at the vicinity of the Fermi
energy, thus substantially influencing the electronic character of the
CNT.  These features are expected to abide up to CNTs lengths of
$60$~$\mu$m, much longer than previously
predicted.\cite{Orlikowski2001}

\begin{figure}[h]
  \begin{center}
    \includegraphics[width=8.60cm]{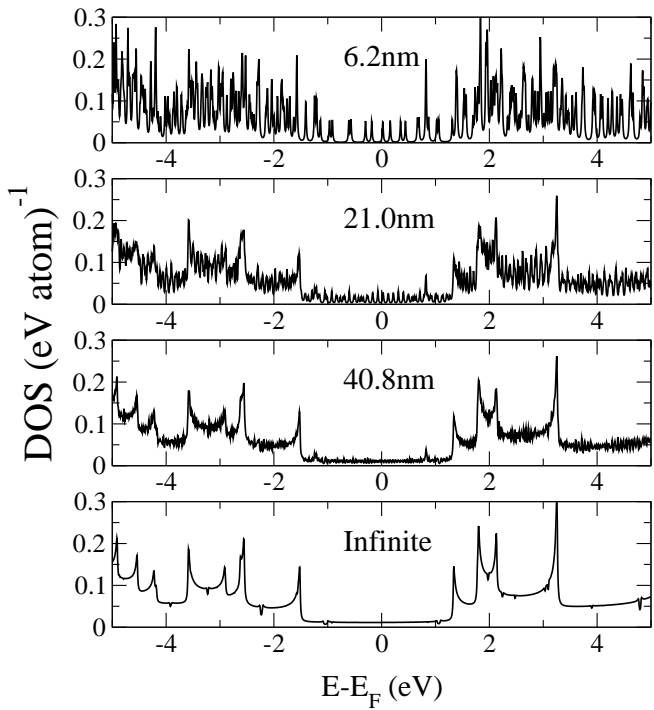}
  \end{center}
  \caption{Density of states of a capped $(6,6)$ CNT calculated for
    several tube lengths and compared to the DOS of the periodic
    system (lower panel).  The energy axis origin of all the panels is
    set to the Fermi energy of the periodic system ($-4.42$~eV)}
  \label{Fig: DOS metallic}
\end{figure}
The relaxed structures of the CNTs employed in this study are
presented in Fig.~\ref{Fig: CNT Geometries}.  These configurations
have been obtained using Pople's Gaussian basis
set~\cite{Hehre1972,BasisSet} and the
Perdew-Burke-Ernzerhof~\cite{Perdew1996,Perdew1997} realization of the
generalized gradient approximation of density functional theory as
implemented in the {\it Gaussian} suite of
programs.~\cite{Frisch2004}.  The DOS calculations has been carried out
using the following relation:
\begin{equation}
  \rho(E)=-\frac{1}{\pi}\Im\{Tr[G^r(E)S]\},
  \label{Eq: DOS}
\end{equation}
where, $S$ is the overlap matrix, $G^r(E)=[\epsilon S-H]^{-1}$ is the
retarded Green's function (GF), $\epsilon=E+i\eta$, $E$ is the energy,
$H$ is the Hamiltonian matrix, and $\eta\rightarrow 0^+$ is a small
imaginary part introduced in order to shift the poles of the GF from
the real axis.  The Hamiltonian matrix is calculated using a divide
and conquer DFT approach.~\cite{Hod2006-2} Within this approach $H$ is
represented by a block tridiagonal matrix, where the first and last
diagonal blocks correspond to the two capping ends of the CNT.  The
remaining diagonal blocks correspond to the central part of the CNT
which is composed of a replicated unit cell.  The two capping ends and
the replicated central part unit cell are chosen to be long enough
such that the block-tridiagonal representation of $H$ (and $S$) is
valid.  All the electronic structure calculations are performed using
the screened exchange hybrid functional of Heyd, Scuseria, and
Ernzerhof~\cite{Heyd2003}, that was found to provide a good
description of the single-particle band energies of both
semi-conducting~\cite{Barone2005_1} and metallic~\cite{Barone2005_2}
CNTs.  The capping end diagonal Hamiltonian blocks and their coupling
to the central part are evaluated via a molecular calculation
involving the two capping ends and one unit cell cut out of the
central part.  We approximate the central part replicated unit cell
blocks and the coupling between two such adjacent blocks to be
constant along the CNT and extract them from a periodic boundary
condition~\cite{Kudin2000} calculation.  The resulting
block-tridiagonal matrix $(\epsilon S-H)$ is then partially inverted,
using an efficient algorithm,~\cite{Godfrin1991} to obtain the
relevant GF blocks needed for the DOS calculation.  For a detailed
description of our divide and conquer approach see
Ref.[\onlinecite{Hod2006-2}].

In Fig.~\ref{Fig: DOS metallic} we present the DOS of the (6,6)
metallic CNT at an energy range of $\pm 5$~eV around the Fermi energy
of the infinite system for several tube lengths.  Qualitatively, it
can be seen that for a $6.2$~nm CNT (upper panel) the DOS is composed
of a set of irregularly spaced energy levels and is totally
uncorrelated with the DOS of the infinite CNT (lower panel).  As the
length of the CNT is increased the agreement between the DOS of the
finite and the periodic systems increases.  At a length of $21$~nm one
can clearly see the buildup of the Van-Hove singularities and the
constant DOS at the vicinity of the Fermi energy.  When the length of
the CNT exceeds $40$~nm, finite size effects become negligible and
most of the physical features appearing in the DOS of the infinite
system are recovered.  A similar behavior is obtained for the DOS of
the metallic (9,0) CNT (not shown).

In order to quantify these results we apply a linear cross-correlation
analysis utilizing Pearson's formula.~\cite{NumericalRecipes} In
short, given two discrete data sets $X_i$ and $Y_i$, where $0 \leq i
\leq M$ one can calculate their cross-correlation as follows:
\begin{equation}
  r=\frac{\sum_{i}\left[\left(X_i-\bar{X}\right)\left(Y_{i}-\bar{Y}\right)\right]}
  {\sqrt{\sum_i\left(X_i-\bar{X}\right)^2}\sqrt{\sum_i\left(Y_{i}-\bar{Y}\right)^2}}.
  \label{Eq:Pearson's Cross-Correlation}
\end{equation}
Here $r$ is Pearson's cross-correlation factor, $\bar{X}$ is the mean
value of the $X_i$ data set, and $\bar{Y}$ is the mean value of the
$Y_i$ data set.  The cross-correlation factor of Eq.~\ref{Eq:Pearson's
Cross-Correlation} is normalized such that $-1 \leq r \leq 1$.  When
$r=1$ the two sets are completely correlated, when $r=-1$ one set is
completely correlated with the inverse of the other set, and when
$r=0$ there is no significant correlation between the two sets.
\begin{figure}[h]
  \begin{center}
    \includegraphics[width=8.5cm]{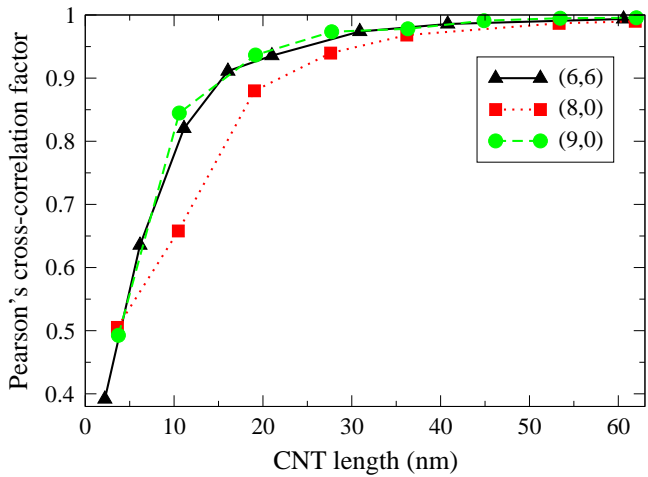}
  \end{center}
  \caption{Pearson's cross-correlation factor between the DOS of
  finite and periodic (6,6) (solid black line), (8,0) (dotted red
  line) and (9,0) (dashed green line) CNTs as a function of the finite
  CNTs lengths.}
  \label{Fig: Cross-Corr}
\end{figure}

In Fig.~\ref{Fig: Cross-Corr} we show the calculated cross-correlation
factor between the DOS of the finite CNTs and that of the infinite
counterparts as a function of the tubes lengths.  For the shortest
CNTs studied, the calculated Pearson's factor indicates no correlation
of the DOS to that of the periodic system.  As discussed above, when
the length of the CNT is increased, the electronic structure of the
infinite system is recovered.  This translates to an asymptotic
Pearson's factor value of $r=1$.

A natural question is: At what length can an infinite metallic CNT be
used in order to represent the electronic structure of its finite
counterpart?  The answer to this question depends, of course, on the
specific type of metallic CNT and on the nature of the electronic
properties considered.  For metallic CNTs one can identify (at least)
two important features in the DOS, namely the {\it Van-Hove
singularities}, which are manifested as characteristic signatures in
the optical spectra of CNTs, and the {\it finite DOS} at the vicinity
of the Fermi energy which accounts for the metallic character of the
CNT.  From a careful analysis of the DOS plots we find that the
Van-Hove singularities are fairly recovered at a length of $\sim
35$~nm for both metallic CNTs considered.  The buildup of a finite DOS
at the vicinity of the Fermi energy is achieved at a slightly larger
length of $\sim 45$~nm.  This is also reflected in the computed
Pearson's cross-correlation factor that is as high as $r=0.98$ at
these CNTs lengths (see Fig.~\ref{Fig: Cross-Corr}).

\begin{figure}[h]
  \vspace{0.5cm}
  \begin{center}
    \includegraphics[width=8.6cm]{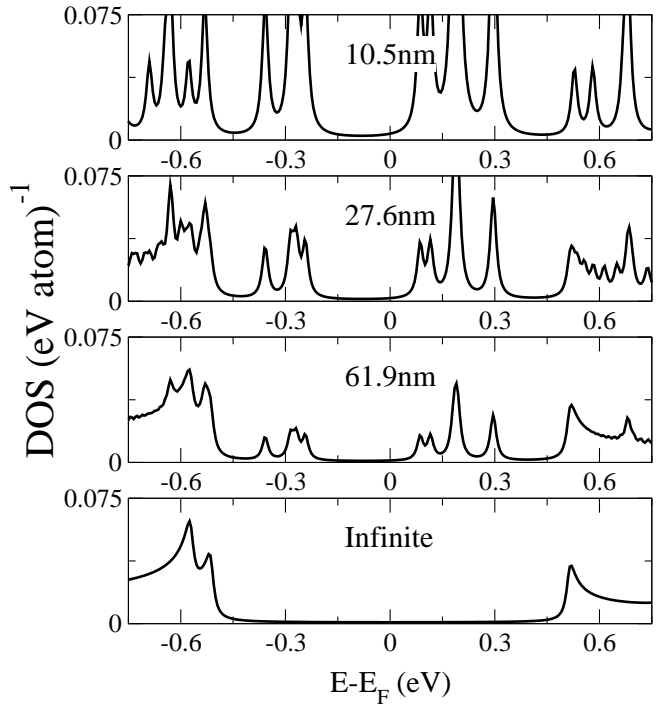}
  \end{center}
  \caption{Density of states of a capped (8,0) CNT calculated for
    several tube lengths and compared to the DOS of the infinite
    periodic system (lower panel).  The energy axis origin of all the
    panels is set to the Fermi energy of the infinite CNT
    ($-4.53$~eV)}
  \label{Fig: DOS semi-conducting}
\end{figure}

We now turn to discuss the semi-conducting CNT.  Similar to the
metallic case, we consider the Van-Hove singularities and the energy
band gap ($E_g$) as two important characteristic features of the its
DOS.  The reconstruction of the Van-Hove singularities as the length
of the finite semi-conducting CNT is increased, follows the same lines
described above for the metallic case.  Nevertheless, a subtle
difference exists.  In Fig.~\ref{Fig: DOS semi-conducting} the DOS of
the finite (8,0) capped CNT is presented for several CNT lengths at a
region of $\pm 0.75$~eV around the Fermi energy of the infinite
periodic system.  While $E_g$ of an infinite periodic (8,0) CNT is
calculated to be $1$~eV (lower panel of Fig.~\ref{Fig: DOS
semi-conducting}), even for the longest CNT considered ($\sim 62$~nm),
a noticeable DOS pattern clearly appears within the energy gap region,
thus reducing the effective gap to a considerably lower value of
$0.25$~eV.  These states are identified as edge effects which
originate from the capping units of the finite CNT.  Their appearance
also translates to a slower approach of the cross-correlation factor
towards its asymptotic value, as can be seen in Fig.~\ref{Fig:
Cross-Corr}.

At this point, it should be mentioned that in order to obtain the DOS
diagrams we have artificially broadened the energy levels by a value
of $1\times10^{-2}$~eV.  In an experimental setup this may correspond
to broadening due to electron-phonon coupling, surface/contacts
effects, and finite temperature effects.  In order for the edge
effects to become negligible it is desired that their peak DOS value
would be considerably smaller than the height of the broadened
Van-Hove singularities corresponding to the top of the valence band
and the bottom of the conduction band.  The height of the edge-related
peaks scales roughly as $\sim 1/L$, where $L$ is the length of the
finite CNT.  For the broadening value mentioned above, a CNT length of
$\sim 6$~$\mu$m is needed for the peaks to become less than $1\%$ in
height with respect to the van-Hove singularity appearing at the
bottom of the conduction band (located $\sim 0.5$~eV above the Fermi
energy).  This value is much larger than the typical CNT length used
in many experimental setups, which implies that the inclusion of
quantum confinement and edge effects can become crucial for the
appropriate interpretation of experimental results.  It should be
noted that the nature of the edge-related DOS peaks appearing within
the energy gap depends on the characteristics of the capping units.
Other systems with different capping units may exhibit different
features.

Edge effects play a dominant role when considering the electronic and
magnetic character of low dimensional elongated systems.  This has
been shown for systems such as graphene
nanoribbons,~\cite{Nakada1996,Fujita1996,Miyamoto1999,Barone2006}
carbon nanotubes,~\cite{Anantram1998, Wu2000, Orlikowski2001,
Compernolle2003, Okada2003, Chen2004-2, Li2005, Jiang2005, Chen2005,
Chen2006, Nemec2006} and other related structures.~\cite{Okada2001}
When considering the study of carbon nanotubes as candidates for
future nano-electronic devices, it is important to identify the
contribution of finite size effects to the physical properties of the
entire system.  In the present letter, we have shown that the limit at
which finite-size effects become negligible, highly depends on the
system under consideration, the capping units, and the physical
property of interest.  This has been done by studying the DOS of CNTs
as a function of their length and comparing with that of the periodic
system.  Our results suggest that for the metallic systems studied, a
CNT segment of, at least, $40-60$nm has to be considered if edge
effects are to be neglected.  For the semi-conducting CNT studied
here, prominent edge effects are expected to dominate up to lengths as
high as a few micrometers.  Even though the exact details of such
effects are expected to depend on the nature of the capping units,
care should be taken regarding their influence on the electronic
character of the system.  Our conclusions should hold for other finite
elongated systems such as graphene nanoribbons, conducting polymers,
and DNA molecules, as well.  More research along these lines is
currently in progress.

This research was supported by the National Science Foundation under
Grant CHE-0457030 and the Welch foundation.  O.H. would like to thank
the generous financial support of the Rothschild and Fulbright
foundations.  Part of the computational time employed in this work was
provided by the Rice Terascale Cluster funded by NSF under Grant
EIA-0216467, Intel, and HP, and by the ADA cluster that is supported
by a Major Research Infrastructure grant from the National Science
Foundation (CNS-0421109), Rice University and partnerships with AMD
and Cray.

\bibliographystyle{./prsty.bst} \bibliography{CNT-DOS}

\begin{thebibliography}{10}
\bibitem{BasisSet}
We use the 6-31G* basis set, which consists of 10$s$, 4$p$ and 1$d$ Gaussian
  functions contracted to 3$s$, 2$p$, and 1$d$.
\end{thebibliography}

\newpage
\end{document}